\documentclass[twocolumn,showkeys,aps,prb,showpacs]{revtex4-1}
\UseRawInputEncoding
\usepackage{graphicx}
\usepackage[CJKbookmarks,dvipdfm,colorlinks,linkcolor=blue,citecolor=blue]{hyperref}
\usepackage{amsmath, amssymb}
\usepackage{makecell}
\usepackage{multirow}
\usepackage{CJK}
\usepackage{mathrsfs}
\usepackage{bm}
\usepackage{amsmath}
\usepackage{dcolumn}
\usepackage{epstopdf}
\usepackage{dsfont}
\usepackage{amssymb}
\usepackage{tabularx}
\usepackage{array}
\usepackage{float}
\usepackage{color}
\usepackage{epstopdf}
\usepackage{mathrsfs}
\usepackage{extarrows}

\begin{document}

\title{Spin-splitting in electric-potential-difference antiferromagnetism}

\author{San-Dong Guo}
\affiliation{School of Electronic Engineering, Xi'an University of Posts and Telecommunications, Xi'an 710121, China}
\begin{abstract}
The  antiferromagnetic (AFM)  materials  are robust to external
magnetic perturbation due to missing any net magnetic moment. In general, the spin splitting  in the band
structures disappears in these antiferromagnets. However, the altermagnetism can achieve spin-split bands in collinear symmetry-compensated antiferromagnet with special magnetic space group. Here,  we propose a new
mechanism that can  achieve spin splitting in two-dimensional (2D) Janus A-type
AFM materials.  Since the built-in electric field caused by Janus structure creates a layer-dependent electrostatic potential, the electronic bands in different layers will stagger,  producing the spin splitting, which  can be called electric-potential-difference antiferromagnetism (EPD-AFM).
 We demonstrate that Janus monolayer $\mathrm{Mn_2ClF}$  is a possible candidate to achieve the EPD-AFM by the first-principles calculations.
It is proposed that the spin splitting can be tuned in EPD-AFM by piezoelectric effect.
 Our works provide a new  design principle for generating  spin
polarization in 2D AFM  materials.

\end{abstract}
\keywords{Spin-split bands, Antiferromagnetism, Built-in electric field~~~~~~~~~~Email:sandongyuwang@163.com}

\maketitle

\section{Introduction}
The spin splitting  in the band structures can be produced  by
utilizing the effect of spin-orbit coupling (SOC)\cite{gs1}.
A general form of the SOC Hamiltonian $H_{SOC}$  in solid-state materials with a lack of inversion symmetry can be expressed as\cite{gs2,gs3}:
 \begin{equation}\label{gs}
    H_{SOC}=\vec{\Omega}(\vec{k})\cdot\vec{\sigma}=\alpha(\vec{E}\times\vec{k})\cdot\vec{\sigma}
 \end{equation}
Where the $\vec{\Omega}(\vec{k})$ is known as a spin-orbit field (SOF) as an effective magnetic field,
$\alpha$ is the strength of the SOC, $\vec{E}$ is the local electric field induced
by the crystal inversion asymmetry, $\vec{k}$ is is the wave vector, and $\vec{\sigma}$=($\sigma_x$, $\sigma_y$, $\sigma_z$) are the Pauli matrices.

If a two-dimensional (2D) material possesses out-of-plane built-in electric field, \autoref{gs} will become:
\begin{equation}\label{h11}
    H_{SOC}=\alpha_R(k_x\sigma_y-k_y\sigma_x)
 \end{equation}
This is known as  Rashba SOC  Hamiltonian\cite{gs4}, and $\alpha_R$ is the so-called Rashba parameter. Here,  the spin
$S$ only has the in-plane components $S_x$ and $S_y$, which depend on the momentum of electrons.
 The impurities and defects can change the momentum of electrons, which can
randomize the spin due to the $k$-dependent SOF, and then induce spin decoherence through the Dyakonov-Perel (DP) mechanism\cite{gs5}.

If a 2D material possesses in-plane built-in electric field, for example along $x$ direction, \autoref{gs} will be reduced into:
\begin{equation}\label{gss}
    H_{SOC}=\alpha_Dk_y\sigma_z
 \end{equation}
 Here, the  spin $S$ only has the out-of-plane component $S_z$. The SOF orientation of \autoref{gss} is
unidirectional,  which  will
lead to a spatially periodic mode of the spin polarization,  known as the persistent spin helix (PSH)\cite{p7,p8}. The PSH can suppress spin dephasing  due to SU(2) spin rotation symmetry,  producing an
extremely long spin lifetime\cite{p7,p9}.

\begin{figure*}
  \includegraphics[width=13cm]{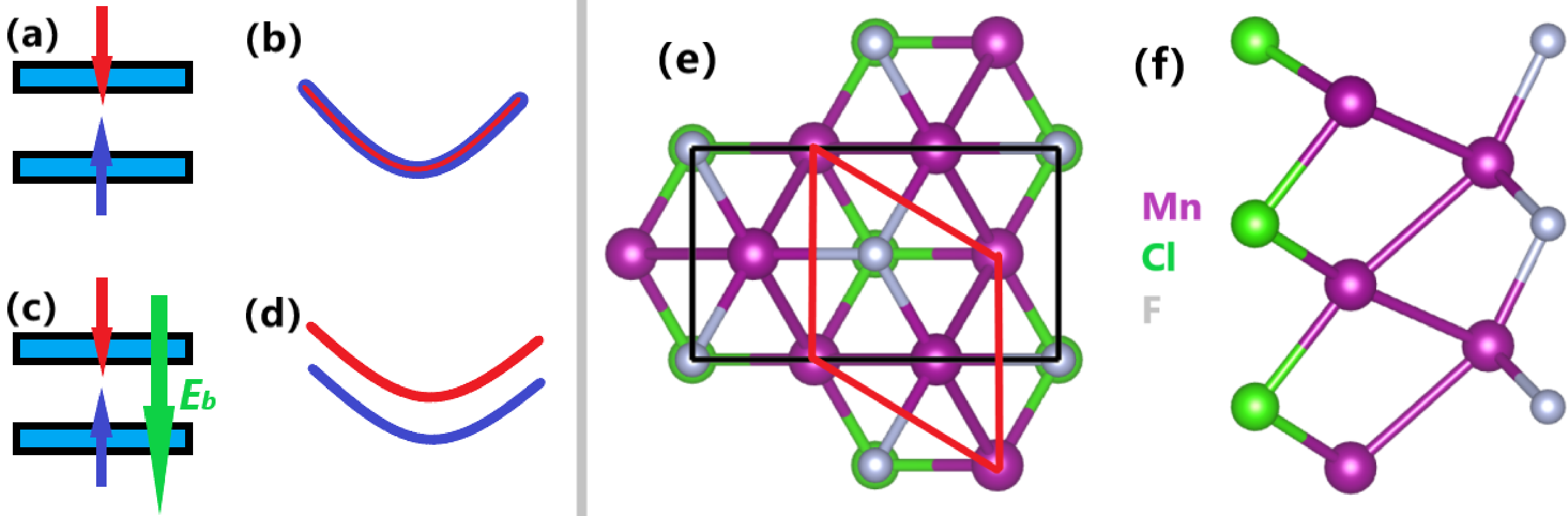}
  \caption{(Color online) (a): for a 2D material, the magnetic atoms have opposite layer spin polarization (A-type antiferromagnetic ordering) without the out-of-plane built-in electric field, producing the degeneration of electron spin (b); (c): for a 2D Janus  material, the magnetic atoms have opposite layer spin polarization (A-type antiferromagnetic ordering) with the out-of-plane built-in electric field $E_b$, destroying the degeneration of electron spin (d). (e) and (f): for Janus monolayer $\mathrm{Mn_2ClF}$, the top and side views of the  crystal structures. In (a), the rhombus primitive cell (rectangle supercell) is marked by the red (black) frame.}\label{st}
\end{figure*}

The spin splitting  can also be observed  in ferromagnetic
(FM) materials.  Superior to FM materials, the  antiferromagnetic (AFM)  materials  are robust to external
magnetic perturbation due to missing any net magnetic moment,  which allows high-speed device operation\cite{k1,k2}.
In general, the spin splitting  in the band
structures is lacking in these antiferromagnets.
However,  the spin splitting has been  realized  in collinear symmetry-compensated antiferromagnet, and  the SOC  is  not needed,  which is called
altermagnetism\cite{k4,k5,k6}.  Several 2D  materials
have  been predicted to be altermagnetic materials, such as $\mathrm{Cr_2O_2}$\cite{k11,k12}, $\mathrm{Cr_2SO}$\cite{k12-1}  and  $\mathrm{V_2Se_2O}$\cite{k13}.

 Here,  we propose a new
mechanism to achieve spin splitting in  AFM materials. For a 2D material, the magnetic atoms have opposite layer spin polarization, namely A-type AFM ordering. If the out-of-plane built-in electric field is lacking, the degeneration of electron spin in the band structures is observed (\autoref{st} (a) and (b)). For a 2D Janus  material, the magnetic configuration is still A-type AFM ordering, but it has an out-of-plane built-in electric field $E_b$, which will destroy the degeneration of electron spin in the band structures (\autoref{st} (c) and (d)). This is because  the built-in electric field creates a layer-dependent electrostatic potential, and the electronic bands in different layers will stagger, which  gives rise to the spin splitting. The spin splitting in 2D Janus A-type AFM materials  can be called electric-potential-difference antiferromagnetism (EPD-AFM).

Recently, the electric-field control of spin polarization
in 2D A-type AFM semiconductor $\mathrm{Mn_2Cl_2}$ has been reported, and the 100\% spin polarization  via electric field can be achieved\cite{k14}.
Based on $\mathrm{Mn_2Cl_2}$,  Janus monolayer $\mathrm{Mn_2ClF}$ is constructed by  replacing one of two Cl layers with F atoms, which  is proved to be a possible candidate to achieve the EPD-AFM by the first-principles calculations. Calculated results show that EPD-AFM in $\mathrm{Mn_2ClF}$  is robust against the electronic correlation. The piezoelectric properties of $\mathrm{Mn_2ClF}$ are also investigated, and the out-of-plane  piezoelectric response may be used to tune the spin splitting. These findings enrich the types of spin splitting, which  is useful for spintronic device applications.

\section{Computational detail}
 Within density functional theory (DFT)\cite{1}, the spin-polarized  first-principles calculations are carried out within the projector augmented-wave (PAW) method by using the standard VASP code\cite{pv1,pv2,pv3}. We use the generalized gradient
approximation  of Perdew-Burke-Ernzerhof (PBE-GGA)\cite{pbe}  as the exchange-correlation functional. To account for electron correlation of Mn-3$d$ orbitals, we use a Hubbard correction $U_{eff}$=4.00 eV\cite{u1,u2,u3} within the
rotationally invariant approach proposed by Dudarev et al.
The kinetic energy cutoff  of 500 eV,  total energy  convergence criterion of  $10^{-8}$ eV, and  force convergence criterion of 0.0001 $\mathrm{eV.{\AA}^{-1}}$ are set to obtain the accurate results.
A  vacuum of more than 16 $\mathrm{{\AA}}$ is used to avoid out-of-plane interaction.

The elastic stiffness tensor  $C_{ij}$  and piezoelectric stress tensor $e_{ij}$   are calculated by using strain-stress relationship (SSR) method and density functional perturbation theory (DFPT) method\cite{pv6}, respectively. The  $C^{2D}_{ij}$/$e^{2D}_{ij}$ has been renormalized by   $C^{2D}_{ij}$=$L_z$$C^{3D}_{ij}$/$e^{2D}_{ij}$=$L_z$$e^{3D}_{ij}$, where the $L_z$ is  the length of unit cell along $z$ direction. We use a 21$\times$21$\times$1  k-point meshes to sample the Brillouin zone (BZ) for calculating electronic structures and elastic properties, and  a 10$\times$21$\times$1  k-point meshes for piezoelectric calculations.
 The interatomic force constants (IFCs)  are calculated by using a  5$\times$5$\times$1 supercell within  finite displacement method, and the phonon dispersion spectrum  can be calculated by the  Phonopy code\cite{pv5}. The elastic, piezoelectric, phonon and ab-initio molecular dynamics (AIMD) calculations are all performed with AFM1 magnetic configuration.

\section{Crystal structure and stability}
Monolayer $\mathrm{Mn_2ClF}$  possesses similar crystal structures with $\mathrm{Mn_2Cl_2}$\cite{k14},   consisting of four atomic layers in the sequence of Cl-Mn-Mn-F (see \autoref{st} (e) and (f)). It is clearly seen that the magnetic Mn atoms distribute in two layers, and  an intrinsic polar electric field along the out-of-plane direction can be induced due to the  different electronegativity of the Cl and F elements, which provides possibility to realize EPD-AFM.
The Janus monolayer   $\mathrm{Mn_2ClF}$ can be constructed  by  replacing one of two Cl layers with F atoms in monolayer  $\mathrm{Mn_2Cl_2}$.  The   $\mathrm{Mn_2Cl_2}$  possesses  $P\bar{3}m1$  space group (No.164), and
 the space group of $\mathrm{Mn_2ClF}$  is reduced into  $P3m1$  (No.156) due to broken horizontal mirror symmetry, which will produce both in-plane and out-of-plane piezoelectricity.

 To determine magnetic ground state of $\mathrm{Mn_2ClF}$,  the rectangle supercell (see \autoref{st} (e)) is used to construct
  FM and three  AFM configurations (AFM1, AFM2 and AFM3). These magnetic configurations are shown in FIG.1 of electronic supplementary information (ESI), and the AFM1  is called A-type AFM state. Calculated results show that the AFM1  configuration is ground state of $\mathrm{Mn_2ClF}$, and its energy per unit cell is 0.43 eV, 0.32 eV and  0.23 eV  lower  than those of FM, AFM2 and AFM3 cases by GGA+$U$. The optimized lattice constants  $a$=$b$=3.43 $\mathrm{{\AA}}$ by GGA+$U$ for AFM1 case. The magnetic easy-axis is confirmed  by magnetic anisotropy energy (MAE), which is defined as  the energy
difference of the magnetization orientation along the (100)
and (001) cases within SOC.  The Calculated  MAE is only  1 $\mathrm{\mu eV}$/Mn, which indicates that the easy-axis of $\mathrm{Mn_2ClF}$ is out-of-plane.

To validate the dynamic,  thermal and mechanical stabilities of $\mathrm{Mn_2ClF}$, the
 phonon spectra, AIMD and elastic constants are calculated,  respectively.
The calculated phonon spectrum  of $\mathrm{Mn_2ClF}$  with  no obvious imaginary frequencies is plotted in FIG.2 of ESI,  indicating its dynamic stability. The AIMD  simulations  using NVT ensemble are carried out  for more than
8000 fs with a time step of 1 fs by using a 4$\times$4$\times$1 supercell at 300 K. According to FIG.3 of ESI,
 during the simulation, the crystal structures of  $\mathrm{Mn_2ClF}$ are maintained without structural fracture, and
the energies are kept stable, confirming
its thermal stability.
Two independent elastic constants $C_{11}$ and $C_{12}$ of $\mathrm{Mn_2ClF}$ are 56.66 $\mathrm{Nm^{-1}}$ and 17.22 $\mathrm{Nm^{-1}}$,  which  satisfy the  Born  criteria of mechanical stability:
$C_{11}>0$ and $C_{11}-C_{12}>0$\cite{ela},  confirming  its mechanical stability.

\begin{figure}
  \includegraphics[width=8cm]{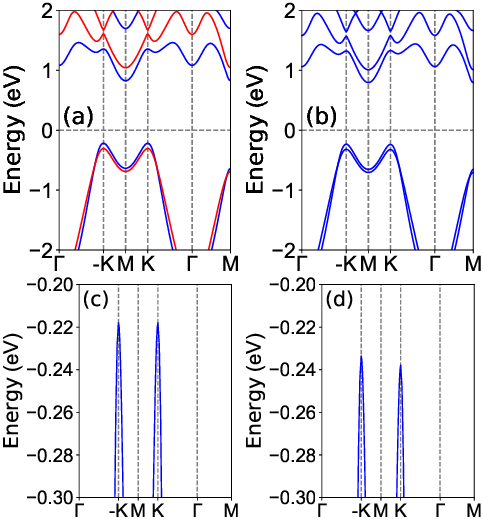}
\caption{(Color online)For  $\mathrm{Mn_2ClF}$,   the energy band structures without SOC (a) and with SOC (b). In (a), the spin-up and spin-down channels are depicted in blue and red. The  (c) and (d) are the partial enlarged drawing  of (a) and (b) near the Fermi level for the valence bands.}\label{band}
\end{figure}

\begin{figure*}
  \includegraphics[width=15cm]{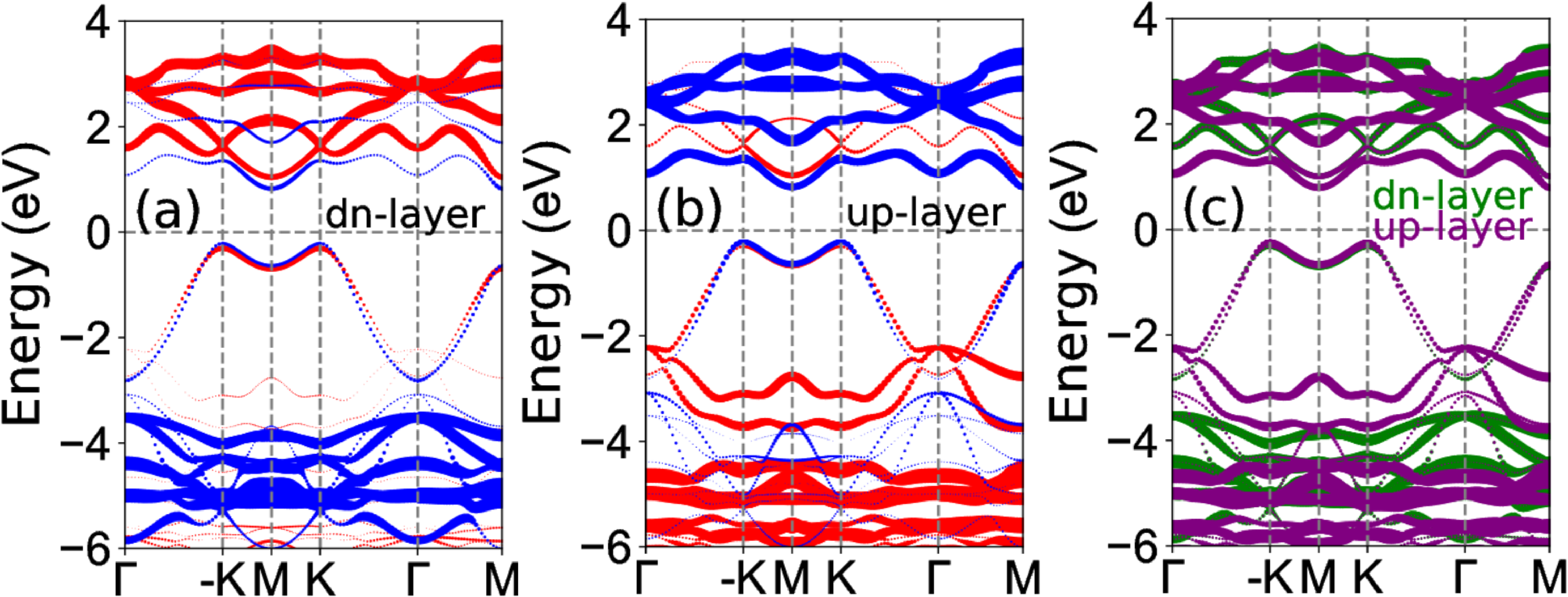}
  \caption{(Color online)For $\mathrm{Mn_2ClF}$, the layer-characters energy band structures without SOC (a and b) and with SOC (c). In (a) and (b), the spin-up and spin-down channels are depicted in blue and red.  The (a) means  dn-layer-characters energy band structures, while the (b) shows up-layer-characters energy band structures. In (c), the spin-up and spin-down channels are not distinguished.  }\label{pro}
\end{figure*}
\begin{figure}
  \includegraphics[width=8cm]{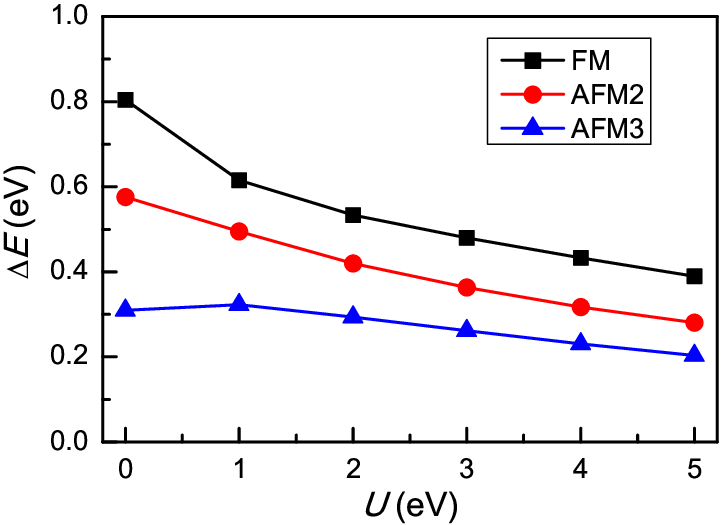}
  \caption{(Color online)For $\mathrm{Mn_2ClF}$, the energy differences between FM/AFM2/AFM3 and AFM1 (per unit cell) as a function of $U$. }\label{ene}
\end{figure}
\begin{figure*}
  \includegraphics[width=12cm]{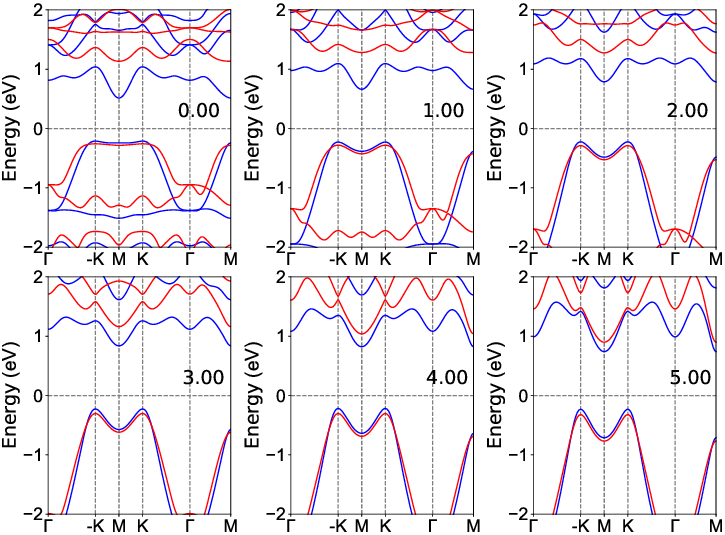}
\caption{(Color online) The energy  band structures of  $\mathrm{Mn_2ClF}$ at representative $U$ without SOC. The spin-up and spin-down channels are depicted in blue and red.}\label{band-u}
\end{figure*}
\section{electronic structures}
The magnetic moments of bottom and top Mn atoms are 4.57 $\mu_B$ and -4.52 $\mu_B$, and total magnetic moment per unit cell is strictly 0.00 $\mu_B$.
In general, no spin splitting can be observed for AFM material. However, our proposed $\mathrm{Mn_2ClF}$ shows obvious spin splitting from calculated energy band structures without SOC in \autoref{band} (a). This is very different from energy band structures of $\mathrm{Mn_2Cl_2}$ (see FIG.4 of ESI), where no spin splitting exists. This difference is because the $\mathrm{Mn_2ClF}$ possesses the out-of-plane polar electric field, while the built-in electric field of $\mathrm{Mn_2Cl_2}$ disappears.
It is clearly seen that   $\mathrm{Mn_2ClF}$ is  an indirect band gap semiconductor with gap value of 1.043 eV.   The valence band maximum (VBM) and conduction band bottom (CBM)  are at high symmetry K/-K and  M points, respectively, and they are provided by the same spin-up channel.
When including the SOC,  the energy band
structures of $\mathrm{Mn_2ClF}$ have very small changes, and it is still an indirect bandgap semiconductor with reduced gap value of 1.028 eV (\autoref{band} (b)).
Without considering SOC, the  K and -K valleys of valence bands are exactly
degenerate (\autoref{band} (c)). However, when SOC is switched on, the energy
degeneracy between the K and - K valleys is lifted  due to broken space- and time-inversion symmetries,  leading to an interesting phenomenon of the
spontaneous valley polarization with very small valley splitting of 4.3 meV (\autoref{band} (d)). This is different from the common valley splitting in FM materials\cite{duan}. Recently, the spontaneous valley
polarization is also predicted in 2D AFM  $\mathrm{Mn_2P_2S_3Se_3}$ with a valley splitting of 16.3 meV\cite{jmc}.
For $\mathrm{Mn_2ClF}$, the layer-characters energy band structures without SOC  and with SOC are plotted in \autoref{pro}.
Calculated results show that the weights of  spin-up and spin-down of both valence and conduction bands are reversed in different Mn layers (\autoref{pro} (a) and (b)), which gives rise to the obvious spin splitting. According to \autoref{pro} (c), it is clearly seen that two Mn layers are non-equivalent due to a layer-dependent electrostatic potential caused by the built-in electric field.

The electronic correlation  can produce important effects on the magnetic ground state, electronic structures and topological properties of 2D magnetic materials\cite{re1,re2,re3,re4,re5}. To confirm robust EPD-AFM, the electronic correlation effects on physical properties  of $\mathrm{Mn_2ClF}$ are considered by using different $U$ values. Firstly, the lattice constants $a$ of $\mathrm{Mn_2ClF}$  are optimized by GGA+$U$ (0-5 eV), and then calculate its related physical properties. Based on FIG.5 of ESI,  the lattice constants $a$  (3.286 $\mathrm{{\AA}}$-3.447 $\mathrm{{\AA}}$) increases with increasing $U$.
To achieve EPD-AFM, the AFM1 magnetic configuration as the ground state of $\mathrm{Mn_2ClF}$ is a crucial factor. So, the energy differences between FM/AFM2/AFM3 and AFM1 (per unit cell) as a function of $U$ are plotted in \autoref{ene}.
 It is found that $\mathrm{Mn_2ClF}$ is always a AFM1 ground state in considered $U$ range.
The evolutions of energy  band
structures as a function of $U$  are plotted in \autoref{band-u}, and the total gap vs $U$ is shown in FIG.6 of ESI.
In considered $U$ range, the $\mathrm{Mn_2ClF}$ is always an indirect gap semiconductor, and shows obvious spin splitting.
 The VBM and CBM  are always at high symmetry K/-K and  M points, which  are provided by the same spin-up channel.
Finally, the MAE as a function of $U$ is plotted in FIG.7 of ESI.  When $U$ is less than about 4.7 eV, the out-of-plane magnetic anisotropy can be  maintained.  These results show that the EPD-AFM of  $\mathrm{Mn_2ClF}$ is robust.

\section{Piezoelectric properties}
 The $\mathrm{Mn_2Cl_2}$ monolayer possesses no piezoelectricity because of  inversion symmetry. However, due to broken horizontal mirror symmetry, the monolayer $\mathrm{Mn_2ClF}$  has both in-plane  and  out-of-plane piezoelectricity. The piezoelectric response of a material can be described by the third-rank piezoelectric stress tensor  $e_{ijk}$ and strain tensor $d_{ijk}$,  which can be expressed as the sum of ionic and electronic contributions:
 \begin{equation}\label{pe0}
 \begin{split}
      e_{ijk}=\frac{\partial P_i}{\partial \varepsilon_{jk}}=e_{ijk}^{elc}+e_{ijk}^{ion}\\
   d_{ijk}=\frac{\partial P_i}{\partial \sigma_{jk}}=d_{ijk}^{elc}+d_{ijk}^{ion}
   \end{split}
 \end{equation}
 In which $P_i$, $\varepsilon_{jk}$ and $\sigma_{jk}$ are polarization vector, strain and stress, respectively.  The superscripts $elc$/$ion$  means electronic/ionic contribution. The  $e_{ijk}^{elc}$ and $d_{ijk}^{elc}$  are  called clamped-ion   piezoelectric coefficients, while the $e_{ijk}$ and $d_{ijk}$ are  called relaxed-ion  piezoelectric coefficients. The $e_{ijk}$ is related with  $d_{ijk}$  by elastic tensor $C_{mnjk}$:
 \begin{equation}\label{h1}
    e_{ijk}=\frac{\partial P_i}{\partial \varepsilon_{jk}}=\frac{\partial P_i}{\partial \sigma_{mn}}.\frac{\partial \sigma_{mn}}{\partial\varepsilon_{jk}}=d_{imn}C_{mnjk}
 \end{equation}

By using  Voigt notation,  when only considering the in-plane strain and stress\cite{yd1,yd2,yd3}, the \autoref{h1} with $P3m1$ symmetry can be reduced into:
  \begin{equation}\label{h2}
  \begin{split}
 \left(
    \begin{array}{ccc}
    e_{11} & -e_{11} & 0 \\
     0 & 0 & -e_{11} \\
      e_{31} & e_{31} & 0 \\
    \end{array}
  \right)
  =\left(
    \begin{array}{ccc}
       d_{11} & -d_{11} & 0 \\
      0 & 0 & -2d_{11} \\
      d_{31} & d_{31} &0 \\
    \end{array}
  \right)\\
  \left(
    \begin{array}{ccc}
        C_{11} & C_{12} & 0 \\
     C_{12} & C_{11} &0 \\
      0 & 0 & (C_{11}-C_{12})/2 \\
    \end{array}
  \right)
  \end{split}
    \end{equation}
  With an imposed  uniaxial in-plane strain,   both in-plane and out-of-plane piezoelectric polarization  can be produced ($e_{11}$/$d_{11}$$\neq$0 and $e_{31}$/$d_{31}$$\neq$0). However, when   a biaxial in-plane strain is applied,  the
in-plane component will disappear($e_{11}$/$d_{11}$=0), but the out-of-plane component still exists ($e_{31}$/$d_{31}$$\neq$0). By solving the \autoref{h2}, the  two independent $d_{11}$ and $d_{31}$ can be derived:
\begin{equation}\label{pe2}
    d_{11}=\frac{e_{11}}{C_{11}-C_{12}}~~~and~~~d_{31}=\frac{e_{31}}{C_{11}+C_{12}}
\end{equation}

The orthorhombic supercell (see  \autoref{st} (e)) as the
computational unit cell is used  to calculate the  $e_{11}$/$e_{31}$ of $\mathrm{Mn_2ClF}$.
The calculated $e_{11}$/$e_{31}$ is -0.745$\times$$10^{-10}$/-0.191$\times$$10^{-10}$ C/m  with ionic part -0.647$\times$$10^{-10}$/0.372$\times$$10^{-10}$ C/m  and electronic part -0.098$\times$$10^{-10}$/-0.563$\times$$10^{-10}$ C/m. For $e_{11}$,   the same signs can be observed for the electronic and ionic contributions, and  the ionic part plays a decisive role.
 However, for $e_{31}$, the electronic and ionic contributions  have  opposite signs, and  the electronic part dominates the  piezoelectricity.
Based on \autoref{pe2}, the calculated  $d_{11}$ and $d_{31}$ of $\mathrm{Mn_2ClF}$ are  -1.89 and -0.26 pm/V, respectively.
 The predicted $|d_{31}|$  is higher than or  compared with those of  other 2D known materials\cite{yd1,yd2,yd3}, which provides possibility to tune spin splitting in $\mathrm{Mn_2ClF}$ by piezoelectric effect.

Electric-field  induced  spin splitting  in $\mathrm{Mn_2Cl_2}$ has been confirmed by the first-principles calculations\cite{k14}. The out-of-plane electric field can tune
the spin splitting in $\mathrm{Mn_2ClF}$. When  a biaxial in-plane strain is imposed,   only out-of-plane $d_{31}$ appears, and an out-of-plane electric field can be induced, which can be used to tune spin splitting in $\mathrm{Mn_2ClF}$.
Piezotronic effect on Rashba spin splitting  in a ZnO/P3HT nanowire array structure has been studied experimentally\cite{ydt}. It is found that the Rashba spin splitting can be effectively tuned by inner-crystal piezo-potential created inside the ZnO nanowires.  So, the coupling between spin splitting and piezoelectric effect  may be observed by  EPD-AFM.

\begin{figure}
  \includegraphics[width=7cm]{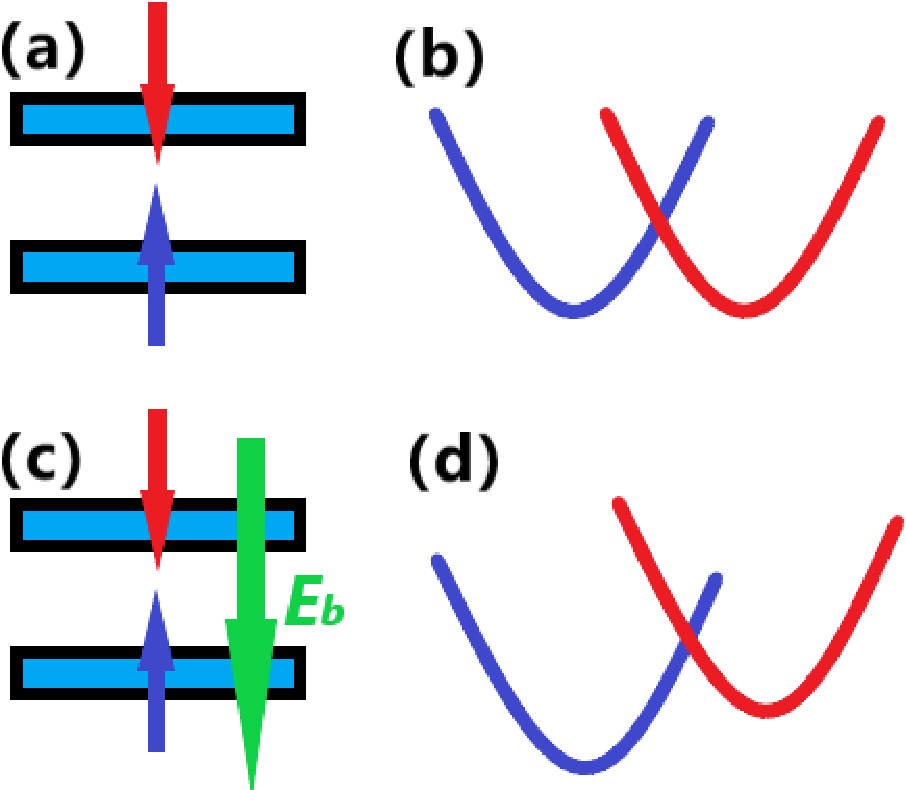}
  \caption{(Color online)(a): for a 2D  altermagnet, the magnetic atoms have opposite layer spin polarization (A-type antiferromagnetic ordering) without the out-of-plane built-in electric field,  destroying the degeneration of electron spin without spin-valley polarization (b); (c): for a 2D Janus  altermagnet, the magnetic atoms have opposite layer spin polarization (A-type antiferromagnetic ordering) with the out-of-plane built-in electric field $E_b$, destroying the degeneration of electron spin with spin-valley polarization (d). }\label{sy}
\end{figure}

\section{Discussion and Conclusion}
 For a 2D altermagnet, the magnetic atoms have opposite layer spin polarization (A-type AFM ordering). If the out-of-plane built-in electric field is lacking, the obvious spin splitting in the band structures can still be  observed (\autoref{sy} (a) and (b)). However, the spin-valley polarization is lacking. Recently, this have been achieved in 2D $\mathrm{Ca(CoN)_2}$\cite{yz}.
 For a 2D Janus  altermagnet, the magnetic configuration is still A-type AFM ordering, but it has an out-of-plane built-in electric field $E_b$, which will produce  spin-valley polarization(\autoref{sy} (c) and (d)). This is because  a layer-dependent electrostatic potential makes electronic bands in different layers stagger, producing  the spin-valley polarization.
 The out-of-plane  polarization filed is equivalent to an external electric field\cite{ar1}. By applying a gate field of 0.2 eV/$\mathrm{{\AA}}$,  monolayer $\mathrm{Ca(CoN)_2}$ possesses a significant spin-valley splitting up to 123 meV\cite{yz}. So, an out-of-plane built-in electric field can induce spin-valley polarization. The 2D Janus A-type altermagnetic material  can be called electric-potential-difference altermagnet (EPD-AM).

In summary,  we propose an alternative strategy to obtain
spin splitting  based on 2D  Janus A-type antiferromagnet.
It is demonstrated that 2D $\mathrm{Mn_2ClF}$ is a possible candidate for realizing EPD-AFM,  which is dynamically, mechanically and thermally  stable.
It is  proved that the EPD-AFM is  robust against electron correlation in $\mathrm{Mn_2ClF}$.
The structural symmetry-breaking  leads to out-of-plane piezoelectric response, providing a possibility to tune spin splitting in $\mathrm{Mn_2ClF}$ by piezoelectric effect. Our works reveal a new 2D family of AFM materials with spin splitting, which allow
high-speed spintronic device applications.

\begin{acknowledgments}
This work is supported by Natural Science Basis Research Plan in Shaanxi Province of China  (2021JM-456).  We are grateful to the  China University of Mining and Technology (CUMT) for VASP software to accomplish this work. We are grateful to Shanxi Supercomputing Center of China, and the calculations were performed on TianHe-2.
\end{acknowledgments}

\end{document}